\documentclass[11pt,twoside]{article}

\usepackage{asp2004}
\usepackage{epsf}
\usepackage{lscape}

\begin{document}
\title{Spitzer Observations of GD 362 and Other Metal-Rich White Dwarfs} 

\author{J. Farihi}
\affil{Gemini Observatory,
	670 North A'ohoku Place, 
	Hilo, HI, USA} 

\author{B. Zuckerman, E. E. Becklin, \& M. Jura}
\affil{Department of Physics \& Astronomy,
	University of California,
	430 Portola Plaza,
	Los Angeles, CA 90095, USA} 
 
\begin{abstract}
A {\em Spitzer} IRAC survey of 17 nearby metal-rich white
dwarfs, nominally DAZ stars, reveals excess emission from 
only 3 targets: G29-38, GD 362 and G167-8.  Observations of
GD 362 with all three {\em Spitzer} instruments reveals a warm
($\approx1000$ K) dust continuum, very strong silicate emission,
and the likely presence of cooler ($\approx500$ K) dust.  While 
there is a general similarity between the mid-infrared spectral
energy distributions of G29-38 and GD 362, the IRAC fluxes of
G167-8 are so far unique among white dwarfs.  However, further
observations of G167-8 are required before the measured excess
can be definitely associated with the white dwarf.

\end{abstract}

\section{Introduction}

The {\em Spitzer Space Telescope} opens up a new phase space
to white dwarf researchers interested in the infrared properties
of degenerate stars and their environments.  The vast majority of
white dwarfs, including the nearest and brightest, are inaccessible
from the ground beyond 2.4 $\mu$m due to their intrinsic faintness
combined with the ever-increasing sky brightness towards longer
wavelengths \citep{gla99}.  This limits any white dwarf science
which aims to study matter radiating at $T<1500$ K.  With the
exception of a few scattered (and, for the most part, unpublished)
ground-based observations of white dwarfs in the relatively benign
10 $\mu$m atmospheric window, only one previously published,
directed mid-infrared study of white dwarfs exists prior to the
launch of {\em Spitzer}; an {\em Infrared Space Observatory}
search for dust emission around 11 nearby white dwarfs, 6 of
which have metal-rich photospheres \citep*{cha99}.

Owing to the superb sensitivity of {\em Spitzer}
\citep{wer04}, a Cycle 1 IRAC program was undertaken to search
for warm dust emission associated with cool hydrogen atmosphere
white dwarfs with photospheric metals, the DAZ stars.  This paper
presents a brief synopsis of the results, including the detection
of $5-8$ $\mu$m flux excess in the IRAC beam of G167-8.  Also
included are new data on GD 362 utilizing all three instruments
aboard the spacecraft.

\section{Scientific Motivation}

The origin of photospheric metals in isolated white dwarfs
has long been an astrophyiscal curiosity \citep*{koe06,koe05,
zuc03,zuc98,dup93a,dup93b,dup92}.  Since any heavy elements
present in the photosphere of a cool white dwarf cannot be
primordial (due to gravitational settling), these metal-rich
degenerates must be externally polluted.  Although accretion
from either the interstellar medium or orbiting dust can, in
priniciple, explain the observed abundances, the relatively
short dwell times for metals in cool hydrogen atmospheres
makes it clear that many of these stars are currently undergoing 
accretion yet are nowhere near known interstellar clouds
\citep{koe06,zuc03}.  It may be the case that both mechanisms
create polluted white dwarf photospheres -- the pristine
nature of a nominally pure hydrogen or helium atmosphere
makes any contamination readily apparent, regardless of the
source.  However, recent developments are beginning to shed 
light on this problem, and there are now a growing number
of metal-rich white dwarfs which are confirmed or suspected
to harbor circumstellar dust \citep*{jur07,far07,kil06,
mul06,jur06,bec05,kil05,jur03}.

\section{Observations}

Between 2004 November and 2005 August, observations of 17
known DAZ white dwarfs were executed with the Infrared Array 
Camera (IRAC; \citealt{faz04}) in all four bandpasses:  3.6, 
4.5, 5.7, \& 7.9 $\mu$m.  A 20 point cycling dither pattern
(medium step size) was used for each target in each bandpass,
with 30 second frame times at each position, yielding a total
exposure time of 600 seconds at all wavelengths.  The data
were processed with the IRAC calibration pipeline (versions
10, 11, \& 12) to create a single, fully-processed and reduced
image upon which to perform photometry.  Aperture photometry
was executed with standard IRAF tasks, and flux measurements
were corrected for aperture size, but not for color.

Additional {\em Spitzer} observations of GD 362 were prompted
by ground-based observations indicating a high probability of
silicate emission around this extremely metal-rich degenerate
\citep{bec05,kil05}.  Spectroscopy over the $5-15$ $\mu$m
region was performed with the Infrared Spectrograph (IRS;
\citealt{hou04}) in 2006 April.  The spectra were taken in 
the two orders of the short-low module, in staring mode, with
a total exposure time of 960 seconds each.  The data were 
processed with IRS calibration pipeline (version 14) and
extracted with the SPICE package.  Observations at 24 $\mu$m
were obtained  with the Multiband Imaging Photometer for Spitzer
(MIPS; \citealt{rie04}) in 2005 September.  The imaging consisted
of 20 cycles with the default 14 point dither pattern and 10
second individual exposures, yielding a total exposure time 
of 2800 seconds.  The data were processed with MIPS calibration 
pipeline (version 12) and aperture photometry was performed with
standard IRAF tasks (including appropriate aperture corrections
for faint sources).

\begin{figure}
\plotone{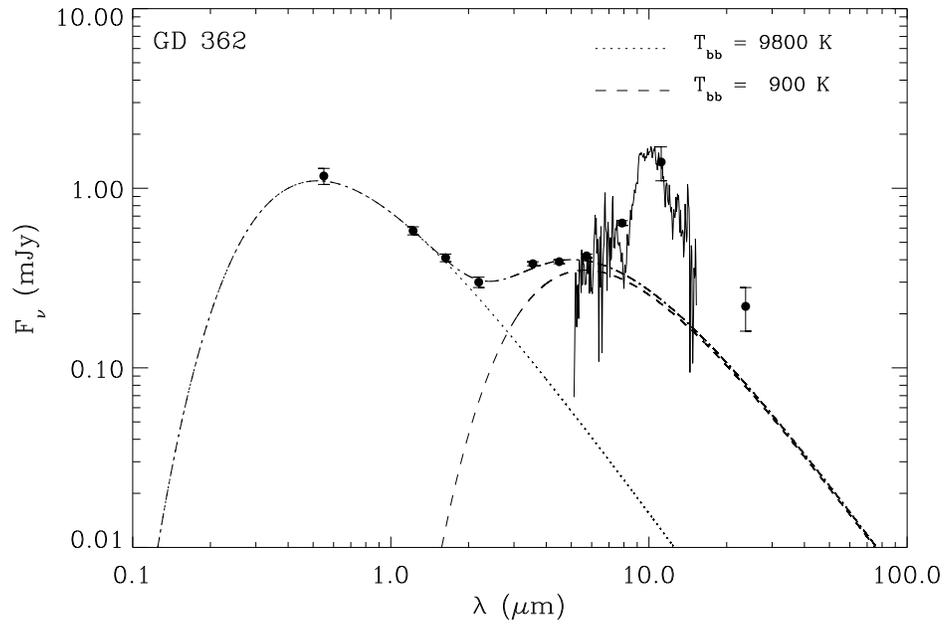}
\caption{Spectral energy distribution of GD 362.  Shown
are data from all three {\em Spitzer} instruments.}
\end{figure}

\begin{figure}
\plotone{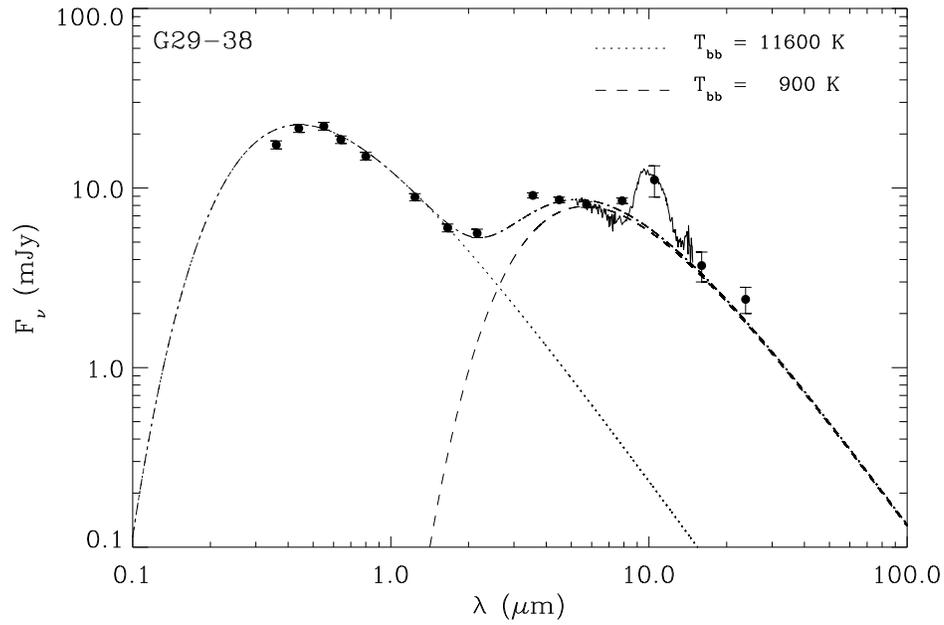}
\caption{Spectral energy distribution of G29-38
for comparison.  Shown are data from all three {\em Spitzer}
instruments.}
\end{figure}

\section{Results}

Of the 17 observed stars, there are 3 which display excess
radiation within their IRAC beams.  The case of G167-8 is
discussed in some detail below, but both G29-38 and GD 362
show warm ($\approx1000$ K) thermal continuum and strong
silicate emission in their IRS spectra and therefore are
confirmed to harbor dust.  Unfortunately, there was a single
DAZ target (G21-16) for which accurate fluxes could not
be extracted due to source confusion -- hence a tentative, 
preliminary statistic is that at least 2 of 16 targets, or
12.5\% have orbiting {\em debris disks}.  Including all 
metal-rich white dwarfs observed with IRAC to date, the
total which display significant {\em flux excess} (this
includes G167-8) is 4 of 25 targets, or 16\% \citep{kil05}.
The full details of the survey, including optical to 
infrared spectral energy distributions, IRAC $3-8$ 
$\mu$m fluxes and uncertainties, are forthcoming
\citep{far07}.

\subsection{GD 362}

The spectral energy distribution of GD 362, together 
with all its {\em Spitzer} data, is displayed in Figure
1.  As expected on the basis of ground-based 10 $\mu$m data
\citep{bec05}, the silicate feature is very prominent -- 
extending even above the peak photospheric flux(!) as well
as contributing significantly to the longest wavelength
IRAC channel ($6.5-9.5$ $\mu$m bandpass, no color correction
applied).  Plotted in Figure 1 specifically are: IRAC $3-8$
$\mu$m photometry, the $5-15$ $\mu$m IRS low resolution
spectrum, and MIPS 24 $\mu$m photometry \citep{jur07,far07}.

For comparison, Figure 2 shows a similar plot for G29-38.
Plotted data are: IRAC $3-8$ $\mu$m photometry \citep{far07};
the $5-15$ $\mu$m IRS low resolution spectrum (downloaded and
extracted from the {\em Spitzer} archive for this work in a
similar manner to the IRS spectrum of GD 362); IRS 16 $\mu$m
peak-up and MIPS 24 $\mu$m photometry \citep{rea05}.  While
the overall similarity (inner dust temperature, silicate
emission) between GD 362 and G29-38 is apparent, there are
some distinctions beyond the relatively much stronger emission
feature of GD 362.  The $3-5$ $\mu$m thermal continuum flux of
GD 362 is rising rather than falling as for G29-38, likely
indicating a difference in the relative amounts of dust cooler
than $\approx1000$ K and perhaps partially contributing to the
strength of the 8 $\mu$m flux of GD 362.  Additionally, the
24 $\mu$m to 2 $\mu$m flux ratio is 0.43 for G29-38 whereas 
for GD 362 the value is nearly twice as large, 0.73.  All of
this certainly indicates more emission from cooler ($\approx500$
K) dust at GD 362 \citep{jur07}.  A complete analysis of the dust
emission from GD 362, including detailed model fits, mass estimates,
temperature distribution, and the sizes and types of the emitting
regions is forthcoming \citep{jur07}.

\subsection{G167-8}

Figure 3 displays the IRAC flux measurements for G167-8 
resulting from aperture photometry.  There are three salient
points regarding this unusual, and so far unique, spectral
energy distribution.  First, there is a source (likely to
be extragalactic) within $6''$ of G167-8 seen in all four
IRAC channels whose flux has been accounted for in all flux
measurements and uncertainties -- the error bars in Figure 3
are quite conservative.  Only at 8 $\mu$m does the flux of
this nearby source become significant, yet its contamination
of the aperture photometry of G167-8 can be accounted for and
removed using a number of straightforward methods.  Second,
it is conceivable that within the $\approx2''$ IRAC beam of 
G167-8 there is another background source contributing flux
at 5.7 and 7.9 $\mu$m.  Ground-based imaging will assist in
evaluating this possibility, yet a final decision may not be
possible without spectroscopy.  Third, G167-8 is a suspected
double degenerate -- as yet unconfirmed \citep{lie05,ber01}.
But a careful look at the $BVRIJHK$ data \citep{zuc03,ber01}
plotted in Figure 3 reveals that a single temperature blackbody
does not fit the data perfectly; in fact, there appears to be
near-infrared excess at $JHK$ (this would become more prominent
if a higher temperature model is used to fit $UBVRI$ only).

Any interpretation of the excess flux of G167-8 warrants 
caution.  If real, the infrared emission at G167-8 would be 
unique among white dwarfs.  All confirmed or suspected debris
disks around metal-rich white dwarfs give rise to measured 
excess flux beginning at $2-3$ $\mu$m \citep{kil06,mul06,
bec05,kil05,rea05}.  If the apparent $5-8$ $\mu$m excess is
confirmed to be associated with G167-8, then some dust disks
may lack warm, opaque debris but harbor only cooler emitting
dust.

\begin{figure}
\plotone{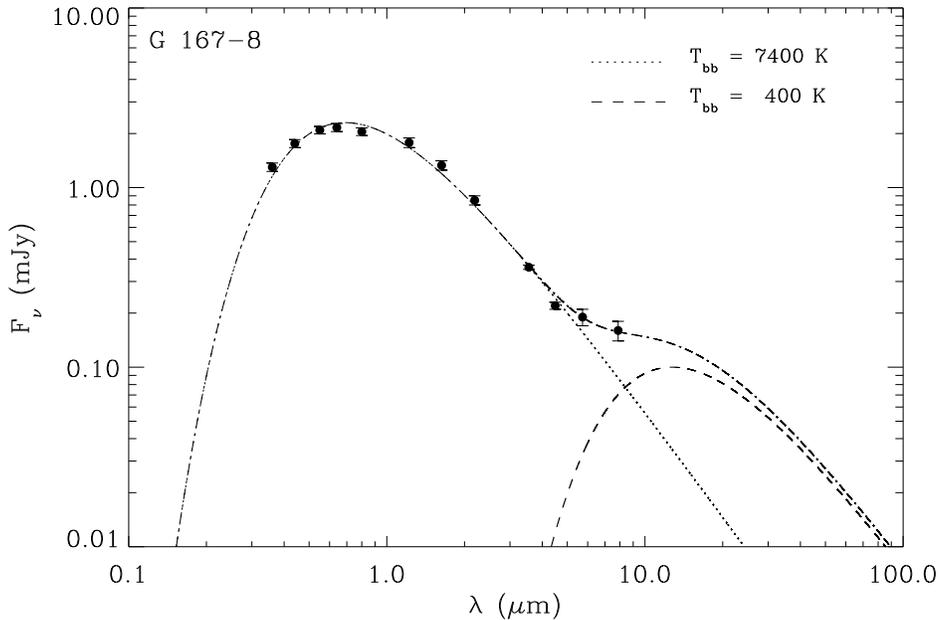}
\caption{Spectral energy distribution of G167-8.}
\end{figure}

\acknowledgements

The authors would like to thanks the conference organizers
for all their hardwork and assistance.  This work is based
on observations made with the Spitzer Space Telescope, which 
is operated by the Jet Propulsion Laboratory, California 
Institute of Technology under NASA contract 1407.

\end{document}